\documentclass[12pt,letterpaper]{article}
\usepackage{jheppub}
\usepackage{graphicx} 
\input{epsf}
\usepackage{epsfig}
\usepackage{epstopdf}
\usepackage{subfigure}

\newcommand{\be}{\begin{equation}}
\newcommand{\ee}{\end{equation}}
\newcommand{\br}{\ensuremath{\Omega}}
\newcommand{\by}{\ensuremath{R}}
\newcommand{\lads}{\ensuremath{L}}
\newcommand{\ket}[1]{\left| #1 \right>}

\title{Null Geodesics, Local CFT Operators and AdS/CFT for Subregions}
\author[a,b]{Raphael Bousso,}
\affiliation[a]{Center for Theoretical Physics and Department of Physics,\\
 University of California, Berkeley, CA 94720, U.S.A.}
\affiliation[b]{Lawrence Berkeley National Laboratory, Berkeley, CA 94720,
  U.S.A.}
  
\author[c]{Ben Freivogel,}
\affiliation[c]{GRAPPA and ITFA, Universiteit van Amsterdam, Science Park 904, 1090 GL Amsterdam, Netherlands}

\author[d]{Stefan Leichenauer,}
\affiliation[d]{California Institute of Technology, Pasadena, CA 91125, U.S.A.}

\author[a,b]{Vladimir Rosenhaus,}

\author[a,b]{and Claire Zukowski}

\abstract{We investigate the nature of the AdS/CFT duality between a subregion
of the bulk and its boundary. In global AdS/CFT in the classical
$G_N=0$ limit, the duality reduces to a boundary value problem that
can be solved by restricting to one-point functions of local operators
in the CFT. We show that the solution of this boundary value problem
depends continuously on the CFT data. In contrast, the AdS-Rindler subregion cannot be continuously reconstructed from local CFT data restricted to the associated boundary region.  Motivated by related results in the mathematics literature, we posit that a
continuous bulk reconstruction is only possible when every null
geodesic in a given bulk subregion has an endpoint on the associated
boundary subregion. This suggests that a subregion duality for
AdS-Rindler, if it exists, must involve nonlocal CFT operators in an
essential way.}

\begin{document}
\maketitle

\section{Introduction}

The AdS/CFT correspondence~\cite{Mal97, Wit98a} provides an important tool for obtaining insight into quantum gravity. Yet even today, the seemingly basic question of how bulk locality is encoded in the boundary theory---in other words, which CFT degrees of freedom describe a given geometrical region in the bulk---has resisted a simple, precise answer. 

In this paper, we investigate the related question of AdS/CFT subregion dualities. That is, we consider the possibility that a CFT restricted to a subset of the full AdS boundary is dual to a geometric subset of the AdS bulk. There is no obvious reason that a geometric region on the boundary has to correspond to a geometric region in the bulk, but there are strong arguments for such a subregion duality in certain simple cases~\cite{CasHue11}, and intriguing hints~\cite{SusWit98, RyuTak06, HubRan07} that it may be true more generally.

The problem of precisely what bulk region should be associated with a
given boundary region is complicated and has been explored recently
by~\cite{BouLei12, CzeKar12, HubRan12}. We will not propose or adopt a rule for constructing such an association. Instead, we focus on one nice feature of the global AdS/CFT duality which does not generalize to arbitrary subregions, namely the ability to reconstruct the bulk using local CFT operators in the classical limit~\cite{HamKab05, HamKab06, Kab11, HeeMar}. Specifically, we will emphasize the role of continuity of the bulk reconstruction, and propose a simple geometric diagnostic testing whether continuous reconstruction holds for a given subregion (see also Ref.~\cite{Porrati:2003na} for related work).

To motivate our investigation, first consider the full global AdS/CFT duality. We will work in Lorentzian signature and fix the Hamiltonian of the CFT, which corresponds to fixing all the non-normalizable modes in the bulk. Now take the $G_N \to 0$ limit in the bulk; the bulk theory reduces to solving classical field equations in a fixed background. The non-normalizable modes are fully determined and non-dynamical, but there are still many allowed solutions because of the normalizable modes. CFT data on the boundary should be sufficient to specify a particular bulk solution. Normalizable modes in the bulk approach zero at the boundary, but a nonzero boundary value can be defined by stripping off a decaying factor,
\be
\phi (b) \equiv \lim_{z \to 0} z^{-\Delta}  \Phi(b,z)~,
\ee
where $z$ is the usual coordinate that approaches zero at the boundary, $b$ stands for the boundary coordinates, and $\Phi$ is a bulk field. We will also use the notation $B = (b,z)$ where convenient.
By the ``extrapolate" version of the AdS/CFT dictionary~\cite{BanDoug98}, these boundary values are dual to expectation values of local operators,
\be
\phi (b) = \langle \mathcal{O} (b) \rangle~.
\ee
We can now ask a classical bulk question: do the boundary values $\phi$ determine the bulk solution everywhere? This is a nonstandard type of Cauchy problem, because we are specifying data on a surface that includes time. 

In a simple toy model where the bulk contains only a single free field with arbitrary mass, Hamilton {\it et al.}~\cite{HamKab05, HamKab06} showed explicitly that this boundary data does specify the bulk solution completely in global AdS. The fact that the boundary data specifies the bulk solution can be considered the classical, non-gravitational limit of AdS/CFT. It is a nontrivial fact that expectation values of local CFT operators are sufficient to reconstruct the bulk field in this case.

\begin{figure}
\centering
\includegraphics[width=4 in]{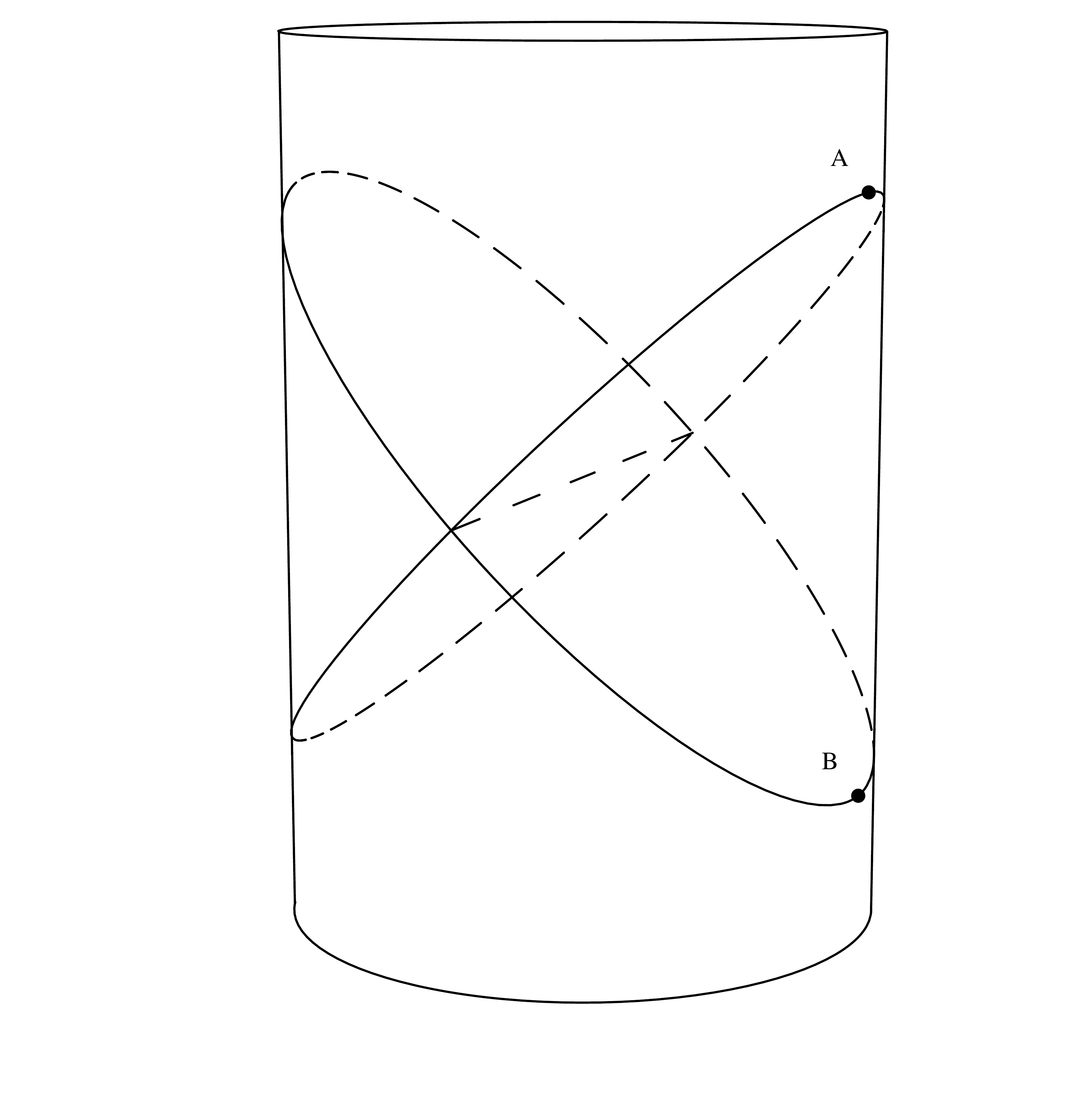}
\caption{Here we show the AdS-Rindler wedge inside of global AdS, which can be defined as the intersection of the past of point $A$ with the future of point $B$. The asymptotic boundary is the small causal diamond defined by points $A$ and $B$. The past lightcone of $A$ and the future lightcone of $B$ intersect along the dashed line, which is a codimension-2 hyperboloid in the bulk. There is a second AdS-Rindler wedge, defined by the points antipodal to $A$ and $B$, that is bounded by the same hyperboloid in the bulk. We refer to such a pair as the ``right" and ``left" AdS-Rindler wedges.}
\label{fig-adsrindler}
\end{figure}

A proposed subregion duality must pass the same test. Is the CFT data in a boundary subregion sufficient to
reconstruct the bulk solution within the corresponding bulk subregion? In
principle, the CFT data is quite complicated. The simplification that occurred in global AdS/CFT, that expectation values of local boundary operators were sufficient, may or may not carry over to other cases, and our task is to properly account for when it does. This is a problem in the theory of classical differential equations which we can hope to solve. Simple examples show that the problem is subtle, however, and to properly capture the physics of the problem we need to differentiate between bulk reconstruction and {\em continuous} bulk reconstruction.

The simplest illustration comes from AdS-Rindler space, which can be described as follows. In the global duality, the CFT is formulated on a sphere cross time and the associated bulk is global AdS. Let us divide the boundary sphere at some time across the
equator. In the bulk, the extremal surface ending on the boundary equator is a
hyperboloid, and we can use
Rindler-type coordinates in AdS so that this extremal surface is a
Rindler horizon. The northern hemisphere on the boundary extends naturally into a small
causal diamond on the boundary, namely the region determined by time
evolution of the data in the northern hemisphere. The corresponding bulk region is a Rindler wedge, shown in Fig.~\ref{fig-adsrindler}, which we will call AdS-Rindler space.

Does the global boundary data, restricted to the small boundary
diamond, determine the bulk solution in the corresponding AdS-Rindler wedge? Hamilton {\it et al.} \cite{HamKab06} also addressed this question. They determined that a particular analytic continuation of the boundary data was necessary to reconstruct the bulk. Here we provide a different answer that does not rely on analytic continuation of the boundary data. We claim that there {\it is} a direct map from the boundary data to the bulk field, but that the map is {\it not continuous}. This leaves the physical interpretation open to doubt.

There are two reasons to focus on the question of
continuity. First, if the subregion duality is correct, we would expect that measuring
boundary data to finite precision should determine the bulk data to a
corresponding precision. This is only true if the bulk solution depends continuously on the boundary data. Second, the question of continuous
reconstruction seems to be mathematically robust; we will be able to
make heuristic contact with nice mathematical theorems about when
continuous reconstruction is possible.

Continuous reconstruction fails because there are finite excitations in the bulk
Rindler wedge with an arbitrarily small imprint on the boundary
data. The physics of these excitations is simple: there exist null
geodesics that pass through the bulk Rindler wedge, but avoid the
boundary diamond. One can construct solutions where geometric
optics is an arbitrarily good approximation and the energy is
concentrated along such a null geodesic. In this way, we can construct
solutions that are finite in the bulk but have arbitrarily small
boundary data in the Rindler wedge.

We can also ask a slightly different mathematical question, which is closely related to bulk reconstruction from the boundary data but simpler to analyze: the question of unique continuation. Suppose we are given the bulk solution in some region near the boundary, and we want to continue the solution further into the bulk. In the AdS context, evolution inward is roughly dual to RG flow in the CFT. This question is closely related to the previous one, and again can be diagnosed with null geodesics \cite{Tat04}. In the case of the bulk Rindler wedge we find that unique continuation fails as well. We cannot evolve the solution radially inward in this case.

Given the connection to continuity and local reconstruction, as well as geometrical simplicity, we are motivated to propose a diagnostic for continuous bulk reconstruction from local CFT operators:
\newline\newline
{\bf Does every null geodesic in the bulk subregion have an endpoint on the corresponding boundary subregion?}
 \newline\newline
\indent Despite the failure of this diagnostic for AdS-Rindler, there are good reasons to think that this particular subregion duality actually holds. The Rindler wedge can be thought of as an eternal black hole with a hyperbolic horizon.  This suggests that a duality holds, by analogy with the ordinary eternal black hole: the CFT in the Hartle-Hawking state may be restricted to one boundary component, and the resulting thermal state is dual to one of the two exterior region of an eternal AdS-Schwarzschild black hole~\cite{CasHue11}. 

 Since continuous reconstruction from CFT one-point functions fails for this subregion, we learn that nonlocal boundary operators must play an important role in the duality even in the classical limit.  Generalizing this result, we learn that nonlocal CFT operators~\cite{PolSus99,Sus99} are important when subregions are small enough that the boundary region no longer captures all null rays passing through the bulk.

The remainder of the paper is organized as follows. In Section~\ref{sec-reconstruction} we review the general procedure for reconstructing the bulk solution from boundary data which was employed by Hamilton {\it et al.} in their work. We also show how to determine continuity of the reconstruction map. The general method is applied to global AdS, AdS-Rindler space, the Poincare patch, and Poincare-Milne space. In Section~\ref{sec-diagnostic} we formulate the geometric diagnostic of capturing null geodesics and relate it to continuity of the reconstruction map, making contact with results in the mathematics literature. We apply the diagnostic to the black hole geometries, as well, without finding an explicit reconstruction map. In Section~\ref{sec-duality}, we exhibit arguments that a subregion duality does exist for AdS-Rindler space. In Section~\ref{sec-disc}, we note that this can be reconciled with the failure of continuous reconstruction from local fields if the duality involves nonlocal boundary operators in an essential way.

\section{The Reconstruction Map}\label{sec-reconstruction}

\subsection{General Formulas}

We begin this section by reviewing the procedure for obtaining a bulk solution from boundary data using eigenmodes of the wave equation, generalizing the approach of Ref.~\cite{HamKab06}. A classical, free bulk field $\Phi$ can be expanded in terms of orthonormal modes $F_k$ which depend on a collection of conserved quantities $k$:
\be
\Phi(B) = \int dk~ a_kF_k(B) + \text{c.c.}~ \label{phiB}
\ee
Near the boundary, the modes $F_k$ have the asymptotic form $F_k(B) \sim r^{-\Delta}f_k(b)$. Thus we find the boundary field $\phi = \lim_{r\to\infty} r^{\Delta}\Phi$ has the expansion
\be
\phi(b)  = \int dk~ a_k f_k(b) + \text{ c.c.}~\label{phib}
\ee
Given $\phi(b)$, we can ask whether it is possible to reconstruct $\Phi(B)$. Recall that $\phi(b)$ is dual to a one-point function in the CFT, hence this is equivalent to asking whether the bulk field is determined by CFT one-point functions. This is possible when the $a_k$ can be extracted from $\phi$ through an inner product of the form
\be\label{bdryFourier}
a_k = W_k \int db~ f^*_k(b) \phi(b)~,
\ee
where $W_k$ is a weighting factor. Equivalently, the boundary mode functions should satisfy the orthogonality relation
\be\label{bdryProduct}
\int db~f_k^*(b)f_{k'}(b) =W_k^{-1} \delta_{k,k'}~.
\ee
There is no guarantee that a relation such as (\ref{bdryProduct}) will hold in general. We will see both possibilities in the examples below.

Given (\ref{bdryFourier}), it is a simple matter to solve for $\Phi(B)$:
\be\label{reconstruction}
\Phi(B) = \int dk~ \left[W_k \int db~ f^*_k(b) \phi(b)\right]F_k(B) + \text{c.c.}~
\ee
We emphasize that at this stage~\eqref{reconstruction} is, in principle, a recipe for computing the bulk field in terms of the boundary field. 

However, there is an important simplification when the order of integration over $k$ and $b$ can be exchanged. Then we have
\be\label{smearing}
\Phi(B) = \int db~K(B|b)\phi(b)~,
\ee
where
\be\label{smearingFunction}
K(B|b) = \int dk~ W_k f_k^*(b) F_k(B) + \text{c.c.}~
\ee
This is a nontrivial simplification which does not occur in all cases. We will see below that when the order of integration is illegitimately exchanged, as in the example of the AdS-Rindler wedge, the integral over $k$ in (\ref{smearingFunction}) does not converge \cite{HamKab06}.\footnote{With certain extra assumptions on the fields, however, \cite{HamKab06} is able to construct a complexified smearing function.}

Non-convergence of the integral in (\ref{smearingFunction}) is due to growth of the eigenmodes at large $k$. The large $k$ behavior of the modes is closely related to the question of continuity of the reconstruction map, $\phi(b) \mapsto \Phi(B)$. To examine continuity, we need to adopt definitions for the bulk and boundary norms. On the boundary, we will follow Ref.~\cite{BarLeb92} and use the norm
\be\label{bdryNorm}
||\phi||_b^2 = \int db~|\nabla_b \phi|^2 + |\phi|^2 ~.
\ee
Here $|\nabla_b\phi|^2$ is positive-definite, not Lorentzian, even though we are in a Lorentzian space-time. In other words, the norm looks like an integral of an energy density (over both space and time), not an action. We will leave its exact form unspecified here, but will be explicit in the examples below. The correct norm to choose is an open question, and a different choice may affect the answer. Our choice is motivated by related results in the mathematics literature, but it may not be a natural choice for this problem. For now, this norm will serve to illustrate the possible answers to the continuity question. Because of (\ref{bdryProduct}), we will find that $||\phi||_b^2 \propto \int p(k)W_k^{-1}|a_k|^2$, where $p(k)$ is a quadratic polynomial in the conserved momenta.

In the bulk, a convenient and natural norm is given by the energy of the solution. Adopting the standard Klein-Gordon normalization for the modes $F_k(B)$, the energy is given by 
\be\label{bulkNorm}
||\Phi||_B^2 =  E[\Phi] =\int dk~|\omega(k)| |a_k|^2~,
\ee
where $\omega(k)$ is the frequency written as a function of the conserved quantities (one of which may be the frequency itself). The reconstruction map is continuous if and only if there is a constant $C>0$ such that
\be
||\Phi||^2_B  \leq C ||\phi||^2_b~.
\ee
That is, a bulk solution of fixed energy cannot have arbitrarily small imprint on the boundary. Equivalently, by going to momentum space, the product $\omega(k)W_k/p(k)$ must be bounded from above. In the remainder of this section we apply these general formulas to several specific cases to find smearing functions and check continuity. We restrict ourselves to a 2+1-dimensional bulk for simplicity.

\subsection{Global AdS}

The AdS$_{2+1}$ metric in global coordinates is
\begin{equation}
ds^2 = -\frac{1}{\cos^2 \rho} dt^2 + \frac{1}{\cos^2\rho}d\rho^2 + \tan^2 \rho\, d\theta^2~.
\end{equation}
The Klein-Gordon equation in these coordinates reads
\begin{equation}
- \cos^2\rho\, \partial^2_t \Phi +  \frac{\cos^2 \rho}{\tan \rho}\partial_\rho \left( \tan\rho \,\partial_\rho \Phi  \right) + \frac{1}{\tan^2 \rho} \partial^2_\theta \Phi = m^2 \Phi~.
\end{equation}
The normalizable solutions are
\be
F_{n l} = N_{nl} e^{-i\omega t}e^{il\theta}\cos^\Delta \rho \sin^{|l|} \rho  \mathcal{F}_{nl}(\rho)\\
\ee
where 
\be
N_{nl} = \sqrt{\frac{\Gamma(n+|l|+1)\Gamma(\Delta + n+ |l|)}{n!\Gamma^2(|l|+1)\Gamma(\Delta+n )}}~,
\ee
\be
 \mathcal{F}_{nl}(\rho) = \,_2F_1(-n,\Delta + n + |l|, |l|+1,\sin^2\rho)~,
\ee
and the frequency is $\omega = \Delta + 2n + |l|$.
The boundary modes are
\be
f_{nl}  = \lim_{\rho \to \pi/2} \cos(\rho)^{-\Delta} F_{nl} = (-1)^n e^{il\theta-i\omega t}\sqrt{\frac{\Gamma(\Delta + n + |l|)\Gamma(\Delta+n)}{n!\Gamma^2(\Delta)\Gamma(n+ |l| + 1)}}~.
\ee

Following the general procedure outlined above, we can compute the smearing function
\be
K(\theta, t,\rho|\theta',t')= \sum_{n,l}~\frac{1}{4\pi^2}\frac{\Gamma(\Delta)\Gamma(n+ |l| + 1)}{\Gamma(\Delta+n) }(-1)^n  e^{-i\omega (t-t')}e^{il(\theta-\theta')}\cos^\Delta \rho \sin^{|l|} \rho  \mathcal{F}_{nl}(\rho) + \text{c.c.}~
\ee
This can be summed to obtain the result of Ref.~\cite{HamKab06}.

The boundary norm in this case is given by\footnote{In global coordinates, the norm in position space is properly defined as an average over time. This is related to the fact that the frequencies are discrete.}
\be
\lim_{T\to \infty} \frac{1}{T}\int_{-T/2}^{T/2} dt d\theta \left((\partial_t\phi)^2 +(\partial_\theta\phi)^2 +\phi^2\right) = 4\pi^2\sum_{nl}(\omega^2 + l^2 +1)\frac{\Gamma(\Delta + n + |l|)\Gamma(\Delta+n)}{n!\Gamma^2(\Delta)\Gamma(n+ |l| + 1)}|a_{nl}|^2~.
\ee
The reconstruction map is continuous if and only if the following quantity is bounded:
\be
\frac{\omega W_{nl}}{1+\omega^2 + l^2} = \frac{\omega}{4\pi^2(1+\omega^2+l^2)}\frac{n!\Gamma^2(\Delta)\Gamma(n+ |l| + 1)}{\Gamma(\Delta + n + |l|)\Gamma(\Delta+n)}~.
\ee
This ratio clearly remains finite for all values of $n$ and $l$, thus proving continuity.

\subsection{AdS-Rindler}

We now turn to the AdS-Rindler wedge, which in 2+1 dimensions has the metric
\be \label{Rindlermetric}
ds^2 =  \frac{1}{z^2}\left[-\left(1 - \frac{z^2}{z_0^2}\right)dt^2 +  \frac{dz^2}{1 - \frac{z^2}{z_0^2}} + dx^2\right]~.
\ee
The Rindler horizon is located at $z=z_0$, while the AdS boundary is at $z=0$. The Klein-Gordon equation is
\be
-\frac{z^2}{1-z^2/z_0^2}\partial_t^2\Phi + z^3\partial_z\left(\frac{1}{z}\left(1-\frac{z^2}{z_0^2}\right)\partial_z\Phi\right) + z^2\partial_x^2 \Phi = m^2\Phi ~.
\ee
The normalizable solutions are
\be
F_{\omega k} =  N_{\omega k}e^{-i\omega t}e^{ikx}z^{\Delta}\left(1-\frac{ z^2}{z_0^2}\right)^{-i\hat{\omega}/2}  \,_2F_1\left(\frac{\Delta -i\hat{\omega}-i\hat{k}}{2}, \frac{\Delta -i\hat{\omega} + i\hat{k}}{2},\Delta, \frac{z^2}{z_0^2}\right)
\ee
where $\hat\omega = \omega z_0$,  $\hat k = kz_0$, and
\be
N_{\omega k} = \frac{1}{\sqrt{8\pi^2|\omega|}}\left|\frac{\Gamma(\frac{\Delta + i\hat\omega +i\hat{k}}{2})\Gamma(\frac{\Delta + i\hat\omega -i\hat{k}}{2})}{\Gamma(\Delta)\Gamma(i\hat\omega)}\right|~.
\ee
 The boundary modes are then
\be
f_{\omega k} =\lim_{z\to 0} z^{-\Delta} F_{\omega k}  = N_{\omega k} e^{ikx-i\omega t}~.
\ee
We can attempt to construct the smearing function following Eq.~\ref{smearingFunction}, but, as discussed below that equation, we will find that the integral over $k$ does not converge:
\begin{align}
&K(x,t,z|x',t')=\\&\frac{1}{4\pi^2}\int dkd\omega ~ e^{ik(x-x')}e^{-i\omega(t- t')} z^{\Delta}\left(1-\frac{ z^2}{z_0^2}\right)^{-i\hat{\omega}/2}  \,_2F_1\left(\frac{\Delta -i\hat{\omega}-i\hat{k}}{2}, \frac{\Delta -i\hat{\omega} + i\hat{k}}{2},\Delta, \frac{z^2}{z_0^2}\right)
\\&=\infty~.
\end{align}
This divergence is due to the exponential growth in $k$ of the hypergeometric function when $k\gg \omega$ \cite{HamKab06},
\be
 \,_2F_1\left(\frac{\Delta -i\hat{\omega}-i\hat{k}}{2}, \frac{\Delta -i\hat{\omega} + i\hat{k}}{2},\Delta, \frac{z^2}{z_0^2}\right) \sim \exp[\hat{k} \sin^{-1}(z/z_0)]~.
\ee

The boundary norm is given by
\be
\int dt dx \left((\partial_t\phi)^2 +(\partial_x\phi)^2 +\phi^2\right) = \int d\omega dk~4\pi^2N_{\omega k}^{2}(1+ \omega^2 + k^2)|a_{\omega k}|^2~.
\ee
We see that the ratio which must be bounded in order that continuity hold is
\be
\frac{\omega W_{nl}}{1+\omega^2+k^2} = \frac{2\omega^2}{1+\omega^2+ k^2}\left|\frac{\Gamma(\Delta)\Gamma(i\hat\omega)}{\Gamma(\frac{\Delta + i\hat\omega +i\hat{k}}{2})\Gamma(\frac{\Delta + i\hat\omega -i\hat{k}}{2})}\right|^2~.
\ee
This ratio remains bounded for fixed $k$, but when $k\gg\omega$ it grows like $\exp(\pi \hat{k})$~. So we find both that the
smearing function does not exist and that continuity fails.

\paragraph{Physical Interpretation} In this case, the problem with reconstructing the bulk solution occurs
regardless of the bulk point we are interested in. The discontinuity
can be understood physically. At first, it is surprising that modes
with $\omega < k$ are even allowed; in the Poincare patch, obtained as the $z_0\to \infty$ limit of AdS-Rindler, they are not.\footnote{In Ref.~\cite{SonSta02}, in the context of the BTZ black hole, it is suggested that these modes are connected with finite temperature effects, and the associated exponential factors are interpreted as Boltzmann weights. We consider this to be very suggestive, but have not found a concrete connection to this work.} Near the Rindler horizon
frequency is redshifted relative to its value at infinity, while
momentum is unaffected. So a local excitation with proper frequency
comparable to its proper momentum appears at infinity as a mode with
$\omega < k$. The modes with $\omega < k$ are confined by a potential
barrier that keeps them away from the boundary; for large $k$ the
height of the barrier is proportional to $k^2$. This causes the
boundary data to be suppressed relative to the bulk by a WKB factor
$\exp(- \int \sqrt V) \sim \exp(- \pi k)$.

We have seen that there is no smearing function in this case because a divergence at large momentum prevents us from exchanging the order of integration. To understand the physical meaning of this divergence, we can ask about computing a more physical quantity, which will regulate the divergence. Instead of trying to find an expression for the bulk field at a specified bulk point, consider instead a bulk field smeared with a Gaussian function of some width $\sigma$ in the transverse direction,
\begin{equation}
\Phi_\sigma (t, x, z) \equiv \int dx^\prime \exp \left(- {(x' - x)^2 \over \sigma^2} \right) \Phi(t, x', z)~.
\ee
We only smear in the $x$ direction because the only divergence is in $k$, and we drop various numerical factors and polynomial prefactors that will be unimportant for our conclusion. We will also set $z_0=1$ (which is always possible by an appropriate scaling of coordinates) for the remainder of this section.

The smeared field has a perfectly fine expression in terms of local boundary fields. We can use symmetries to place the bulk point at $t = x = 0$; then
\begin{equation}
\Phi_\sigma (0, 0, z) = \int dt' dx' K_\sigma(0, 0, z | x', t') \phi (x', t')
\end{equation}
with
\begin{align}
K_\sigma&(0, 0, z | x', t') = \nonumber \\
&\int d \omega d k \,e^{\displaystyle i \omega t' - k x' - k^2 \sigma^2}
 \left(1-z^2\right)^{-i\omega/2}  \,_2F_1\left(\frac{\Delta -i\omega-ik}{2}, \frac{\Delta -i\omega + ik}{2},\Delta, z^2\right)~.
 \end{align}
 The important question is the large $k$ behavior of this function. To get a feeling for it, replace the hypergeometric function by its large $k$ limit,
 \begin{equation}
 _2 F_1 \approx g(\omega, \Delta, z) k^{\Delta - 1} \cosh (2 k \theta)
 \end{equation}
 where $\theta$ depends on the distance from the boundary, $\sin \theta = z$, and $g$ a function that does not depend on $k$. We ignore the polynomial prefactor and focus on the exponential dependence. Performing the integral, we get 
 \begin{equation}
 K_\sigma(0, 0, z | x', t') \ ``=" \  \tilde{g}(t, \Delta, z) \exp \left({\theta^2 \over \sigma^2}- {x'^2 \over \sigma^2} - 2i {\theta \over \sigma^2} x' \right)
 \end{equation}  
where the quotation marks indicate that this is only a cartoon of the correct answer that captures the large momentum behavior of the smearing function. Now we can write the smeared bulk field in terms of the boundary values,
\begin{equation}
\Phi_\sigma(0, 0, z) = \int dx' dt' \,  K_\sigma(0, 0, z | x', t') \phi(x', t')~.
\end{equation}

What is the behavior of this function as we localize the bulk field by taking the width small, $\sigma \to 0$? $K_\sigma$ is strongly dependent on $\sigma$: the maximum value of $K_\sigma$ is exponentially large at small $\sigma$, $K_\sigma^{\rm max} = \exp(\theta^2/\sigma^2)$, where again $\theta$ is related to the distance from the horizon, ranging from $\theta= \pi/2$ at the horizon to $\theta = 0$ at the boundary. It varies rapidly, with characteristic wavenumber $\theta/\sigma^2$, and has a width set by $\sigma$. 

The physical length over which the bulk point is smeared is $\sigma_{\rm phys} =  \sigma/z$, and up to an order-one factor we can approximate $\theta \approx z$. Restoring factors of the AdS radius $\lads$, we find that the smeared smearing function $K_\sigma$ is a rapidly oscillating function with maximum value 
\be
K_\sigma^{\rm max} \sim \exp \left( \lads^2 \over \sigma_{\rm phys}^2 \right)~.
\ee
Note that the dependence on the radial location has disappeared upon writing things in terms of the physical size.
Attempting to measure the bulk field at scales smaller than the AdS radius requires exponential precision in the boundary measurement, because we are trying to compute an order-one answer (the bulk field value) by integrating an exponentially large, rapidly oscillating function multiplied by the boundary field value.

We note here an interesting technical feature of this construction. We chose to compute a bulk operator smeared with a Gaussian profile in the transverse direction. Normally, the exact form of a smeared operator is not physically relevant. In particular, we can ask whether it is possible to construct an analogous function $K$ for smeared bulk operators which have smooth but compact support in the transverse direction. Unfortunately this is impossible. In order to overcome the exponential divergence at large $k$ in the mode functions, we had to smear against a bulk profile which dies off at least exponentially fast at large $k$. Such a function is necessarily analytic in $x$, and hence will not have compact support. Therefore we cannot truly localize our smeared bulk operators in the above construction; some residual leaking to infinity is required.

\subsection{Poincare Patch}

The Poincare patch is the canonical example of a subregion duality that works. With our chosen norms, we will find that continuity actually {\em fails} in the Poincare patch, even though a smearing function exists. This suggests that the Poincare patch may already reveal subtleties that we claim exist in the AdS-Rindler case. However, we will see that the nature of the discontinuity is very different from that of the AdS-Rindler wedge. Later, in Section~\ref{sec-diagnostic}, we will argue that this discontinuity may be a harmless relic of our choice of norm, and that a more reliable answer is given by the geometric criterion presented there.

The metric of the Poincare patch is
\begin{equation}
ds^2 = \frac{dz^2 - dt^2 + dx^2}{z^2}~,
\end{equation}
and the Klein-Gordon equation in these coordinates reads
\begin{equation}
- z^2\, \partial^2_t \Phi +  z^{3}\partial_z \left(\frac{1}{z} \partial_z \Phi  \right) + z^2 \partial^2_x \Phi = m^2 \Phi~.
\end{equation}
In this case we label the eigenmodes by $k$ and $q$, with $q>0$. The frequency is given by $\omega = \sqrt{q^2 + k^2}$. Properly normalized, the modes are $F_{q k} = (4\pi \omega)^{-1/2}e^{ikx}   z \sqrt{q}J_{\nu}(qz)$. We have introduced the notation $\nu = \Delta - 1 = \sqrt{1 + m^2}$. Then the boundary modes are
\be
f_{q k} = \lim_{z\to 0} z^{-\Delta}F_{q k} = \frac{q^{\nu+\frac{1}{2}}}{2^\nu\Gamma(\Delta)}\frac{e^{i(kx-\omega t)}}{\sqrt{4\pi\omega}}~.
\ee
The smearing function can easily be computed,
\be
K(x,t,z|x',t') = \int dqdk~\frac{2^{\nu} \Gamma(\Delta)}{4\pi^2 q^{\nu-1}\omega}e^{ik(x-x')}e^{-i\omega(t-t')} z J_{\nu}(qz) + \text{c.c.}~,
\ee
and this matches with the result of Ref.~\cite{HamKab06}.

The boundary norm is
\be
\int dt dx \left((\partial_t\phi)^2 +(\partial_x\phi)^2 +\phi^2\right) = \int dqdk~\frac{\pi q^{2\nu}}{4^\nu\Gamma^2(\Delta)}(1+ \omega^2 + k^2)|a_{qk}|^2~.
\ee
The ratio which must remain bounded for continuity to hold is
\be
\frac{\omega W_{q k}}{1+\omega^2 + k^2} =  \frac{4^{\nu} \Gamma^2(\Delta)\omega}{\pi q^{2\nu}(1+\omega^2 + k^2)}~.
\ee
For large $q,k$ this remains bounded, but as $q\to 0$ it does
not. The physics of the problem is the following. Starting with any solution, we can perform a
conformal transformation that takes
\be
z \to \lambda z, \ \ \ x \to \lambda x, \ \ \ t \to \lambda t~.
\ee
For large $\lambda$, this moves the bulk solution towards the Poincare
horizon and away from the Poincare boundary, resulting in a small boundary imprint. 
Under this scaling, $q \to \lambda^{-1} q$, so it is exactly the small
$q$ behavior above that allows for such an ``invisible'' solution. 

As stated above, we believe that this discontinuity may merely be a problem of the choice of norm. In particular, this is an ``infrared" discontinuity, and the difficulties of the AdS-Rindler wedge were ``ultraviolet" in character. The smearing function seems to be sensitive only to the ultraviolet discontinuities, which suggests that those are more troublesome. Furthermore, in Section~\ref{sec-diagnostic} we will see that the Poincare patch (marginally) passes the geometric test of continuity while the AdS-Rindler wedge clearly fails. A remaining problem for future work to provide a more concrete connection between ``ultraviolet" and ``infrared" discontinuities and the existence or non-existence of a smearing function.

\subsection{Poincare-Milne}\label{sec-poinmilne}

\begin{figure}
\centering
\includegraphics[width=4 in]{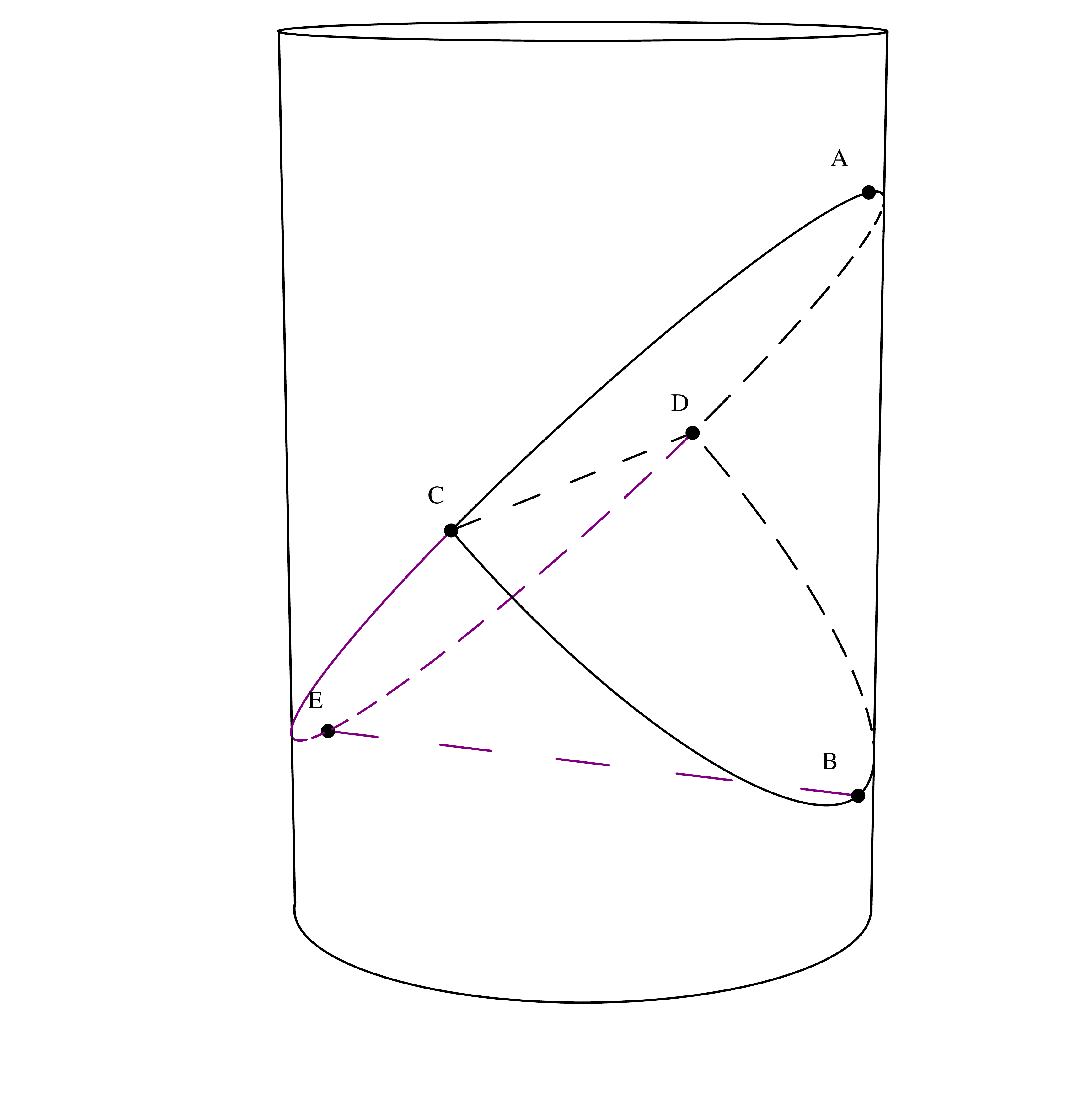}
\caption{Here we depict Poincare-Milne space, together with an AdS-Rindler space that it contains. The bulk of Poincare-Milne can be defined as the intersection of the past of point $A$ with the future of line $BE$. Clearly this region contains the AdS-Rindler space which is the intersection of the past of $A$ and the future of $B$. Furthermore, the asymptotic boundary of the Poincare-Milne space and the AdS-Rindler space is identical, being the causal diamond defined by $A$ and $B$ on the boundary.}
\label{fig-poinmilne}
\end{figure}

Poincare-Milne space is the union of the collection of Milne spaces at each value of $z$ in the Poincare patch. It is useful to contrast the Poincare-Milne case with the AdS-Rindler case considered above. The reason is that the conformal boundary of Poincare-Milne space is identical to that of the AdS-Rindler wedge, but the Poincare-Milne bulk is larger, as shown in Fig.~\ref{fig-poinmilne}.\footnote{For definiteness we discard the future light cone of the point $E$ in the figure, so that the boundary is exactly AdS-Rindler, with no extra null cone.} We expect that the boundary theory of the AdS-Rindler boundary is dual to the AdS-Rindler bulk space and not more~\cite{BouLei12, CzeKar12, HubRan12}, and so it is an important check on our methods that they do not provide false evidence for a Poincare-Milne subregion duality. While we have no proof that the free theory constructions we have considered so far cannot be extended to Poincare-Milne, we can show that the most obvious construction breaks down in a very curious way.

The metric of Poincare-Milne space is
\begin{equation}
ds^2 = \frac{dz^2- dt^2 + t^2dx^2}{z^2}~,
\end{equation}
where we restrict to $t>0$. The Klein-Gordon equation in these coordinates reads
\begin{equation}
- z^2t^{-1}\partial_t(t\partial_t \Phi) +  z^{3}\partial_z \left(\frac{1}{z} \partial_z \Phi  \right) + z^2 t^{-2}\partial^2_x \Phi = m^2 \Phi~.
\end{equation}
The $z$-dependence and $x$-dependence of the normalizable eigenmodes are identical to the Poincare patch case, and the $t$-dependence comes from solving the equation
\begin{equation}
- t^{-1}\partial_t (t\partial_t \psi)  - t^{-2}k^2 \psi  = q^2 \psi~.
\end{equation}
The general solution to this equation is a linear combination of Hankel functions, $\psi  = A H^{(1)}_{ik}(qt)+ B e^{\pi k}H^{(2)}_{ik}(qt) =  A H^{(1)}_{ik}(qt)+ B[H^{(1)}_{ik}(qt)]^* $.

As we will demonstrate, no equation like \ref{bdryProduct} can hold for solutions to this equation. To see this, it is convenient to define $\tilde\psi = (qt)^{1/2} \psi$. Then we have
\begin{equation}
 - \partial^2_t \tilde\psi - \frac{k^2+1/4}{t^{2}}\tilde \psi= q^2 \tilde\psi~.
\end{equation}
This is a Schr\"odinger equation for a scattering state in an attractive $1/t^2$ potential. To simplify the calculation, we will normalize the solutions so that $A=1$ always. The standard expectation from quantum mechanics is that $B$ is then completely determined as a function of $q$, and in particular we will only have a single linearly independent solution for a given value of $q$. However, from the bulk point of view there should always be two solutions for any $q$, corresponding to the positive and negative frequency modes. Indeed, the coefficient $B$ is usually determined by the boundary condition $\tilde\psi(0)=0$, but here that is trivially satisfied for all $B$. Hence $B$ is a free parameter. We will now demonstrate another strange fact about this potential, that eigenmodes with different values of $q$ are not orthogonal, which shows that Eq.~\ref{bdryProduct} does not hold. 

To see this, consider two solutions $\tilde\psi_1$ and $\tilde\psi_2$ corresponding to $q_1$ and $q_2$. We have
\be
(q_1^2 - q_2^2) \int_0^\infty dt ~\tilde \psi_1^* \tilde \psi_2 = \left. \tilde\psi_1^*\partial_t\tilde\psi_2 -\partial_t\tilde\psi_1^* \psi_2 \right|_0^\infty~.\\
\ee
We can compute the inner product once we know the asymptotic behavior of the solutions near $t=\infty$ and $t=0$.

First, we use the large argument asymptotic form of the Hankel function,
\be
H^{(1)}_{ik}(qt) \approx \sqrt{\frac{2}{\pi qt}}e^{i(qt-\pi/4)}e^{k\pi/2}~,
\ee
so that
\be
\tilde \psi_i \approx \sqrt{\frac{2}{\pi}}e^{k\pi/2}\left(e^{i(q_it-\pi/4)}+ B_i e^{-i(q_it-\pi/4)}\right)~.
\ee
Then we find
\be
\lim_{t\to \infty}  \frac{1}{q_1^2 - q_2^2}\left(\tilde\psi_1^* \partial_t \tilde \psi_2 - \partial_t\tilde \psi_1^*\tilde\psi_2\right) = 2e^{\pi k}\left(1+ B_1^*B_2\right)\delta(q_1-q_2)~,
\ee
where we have used the fact that $\lim_{x\to \infty} e^{-iqx}/q = \pi\delta(q)$ and $\delta(q_1+q_2) = 0$ when $q_1$ and $q_2$ are both positive. The result is proportional to a $\delta$-function, as it had to be. For large $t$ the solution approaches a plane wave, and plane waves of different frequencies are orthogonal.

Near $t=0$ we use the small argument expansion
\be
H^{(1)}_{ik}(qt) \approx \frac{1+\coth \pi k}{\Gamma(1+ik)} \left(\frac{qt}{2}\right)^{ik} -\frac{\Gamma(1+ik)}{\pi k} \left(\frac{qt}{2}\right)^{-ik}~,
\ee
so that
\be
\tilde \psi_i \approx C_it^{ik+1/2}+D_it^{-ik+1/2}~,
\ee
where $C_i$ and $D_i$ are determined in terms of $B_i$ and $q_i$. Then we have
\be
\lim_{t\to 0}  \tilde\psi_1^* \partial_t \tilde \psi_2 - \partial_t\tilde \psi_1^*\tilde\psi_2= 2ik\left(C_1^*C_2 - D_1^*D_2\right)~.
\ee
In order to ensure orthogonality, this combination has to vanish for arbitrary choices of the parameters. This is clearly not the case. We note in passing that imposing an extra constraint of the form $D = e^{i\delta}C$, with $\delta$ a new independent parameter, will make the wavefunctions orthogonal. Tracing through the definitions, one can see that this also fully determines $B$ in terms of $q$ and $\delta$, and that $|B|=1$ as expected by unitarity. The choice of $\delta$ corresponds to a choice of self-adjoint extension, necessary to make the quantum mechanics well-defined (for additional discussion of this point see Ref.~\cite{EssGri06}, and see references therein for more on the $1/t^2$ potential in quantum mechanics). As we pointed out above, however, such a prescription is not relevant for our current task, as it would eliminate a bulk degree of freedom.

\subsection{AdS-Rindler Revisited}\label{sec-rindler2}
We would like to emphasize that the above analysis of Poincare-Milne space is not a no-go theorem. As an example, we now show that AdS-Rindler space, analyzed in a certain coordinate system, suffers from the same pathologies. By a change of coordinates, one can show that the AdS-Rindler wedge can be written in a way that is precisely analogous to Poincare-Milne:
\be\label{eq-Rindlermetric2}
ds^2 = \frac{dz^2 - x^2dt^2 + dx^2}{z^2}~,
\ee
where we restrict to the region $x>0$. Using this coordinate system, and following the usual procedure, we encounter problems very similar to those of Poincare-Milne space discussed above. The $x$-dependence of the eigenmodes is found by solving
\be
-x^{-1}\partial_x (x\partial_x\psi) -x^{-2} \omega^2\psi^2  = -q^2\psi~.
\ee
This is equivalent to a Schr\"odinger equation in the same potential as before, except now we are finding bound states instead of scattering states. The analysis is completely analogous to the Poincare-Milne case. There is a continuous spectrum of bound states (unusual for quantum mechanics!), and they are not generically orthogonal. Thus we cannot carry out the program of mapping boundary data to bulk solutions. In this scenario, the choice of a self-adjoint extension would involve quantizing the allowed values of $q$, and by restricting $q$ correctly we can find a set of orthogonal states. While that is appropriate for quantum mechanics, here the bulk physics is well-defined without such a restriction.

\section{A Simple, General Criterion for Continuous Classical Reconstruction:
  Capturing Null Geodesics}\label{sec-diagnostic}

In this section, we propose a general, geometric criterion for classical reconstruction of the bulk from the boundary. To our knowledge, the case of AdS has not been analyzed
explicitly. However, mathematicians such as Bardos {\it et al.} \cite{BarLeb92}
have analyzed the analogous situation in flat spacetime: Consider a
field that solves the classical wave equation in some region \br\ of
Minkowski space with a timelike boundary $\partial\br$, with Neumann boundary
conditions everywhere on the boundary. Now suppose the boundary value
of the field is given in some region $\by \subset \partial\br$ of the boundary. When is
this sufficient to determine the bulk field everywhere in \br?

The central result is that every null geodesic in \br\ should intersect \by\ in order for continuous reconstruction to be possible. The basic
intuition is that if there is some null geodesic that does not hit
\by, then by going to the geometric optics limit we can construct
solutions that are arbitrarily well localized along that
geodesic. These solutions are ``invisible" to the boundary observer
who only can observe $\phi$ in the region \by, in the sense that the boundary imprint can be made arbitrarily small while keeping the energy fixed.

It is not surprising that capturing every null geodesic is a necessary
condition for continuous reconstruction, and this will be the
important point for us. In many situations, however, the null geodesic
criterion is actually sufficient. As long as every null geodesic hits
\by, the entire bulk solution can be reconstructed. (The theorems are
quite a bit more general than we have described here, applying to
general second-order hyperbolic partial differential equations, and
generalizing to nonlinear problems.)

\begin{figure}
\centering
\includegraphics[width=4 in]{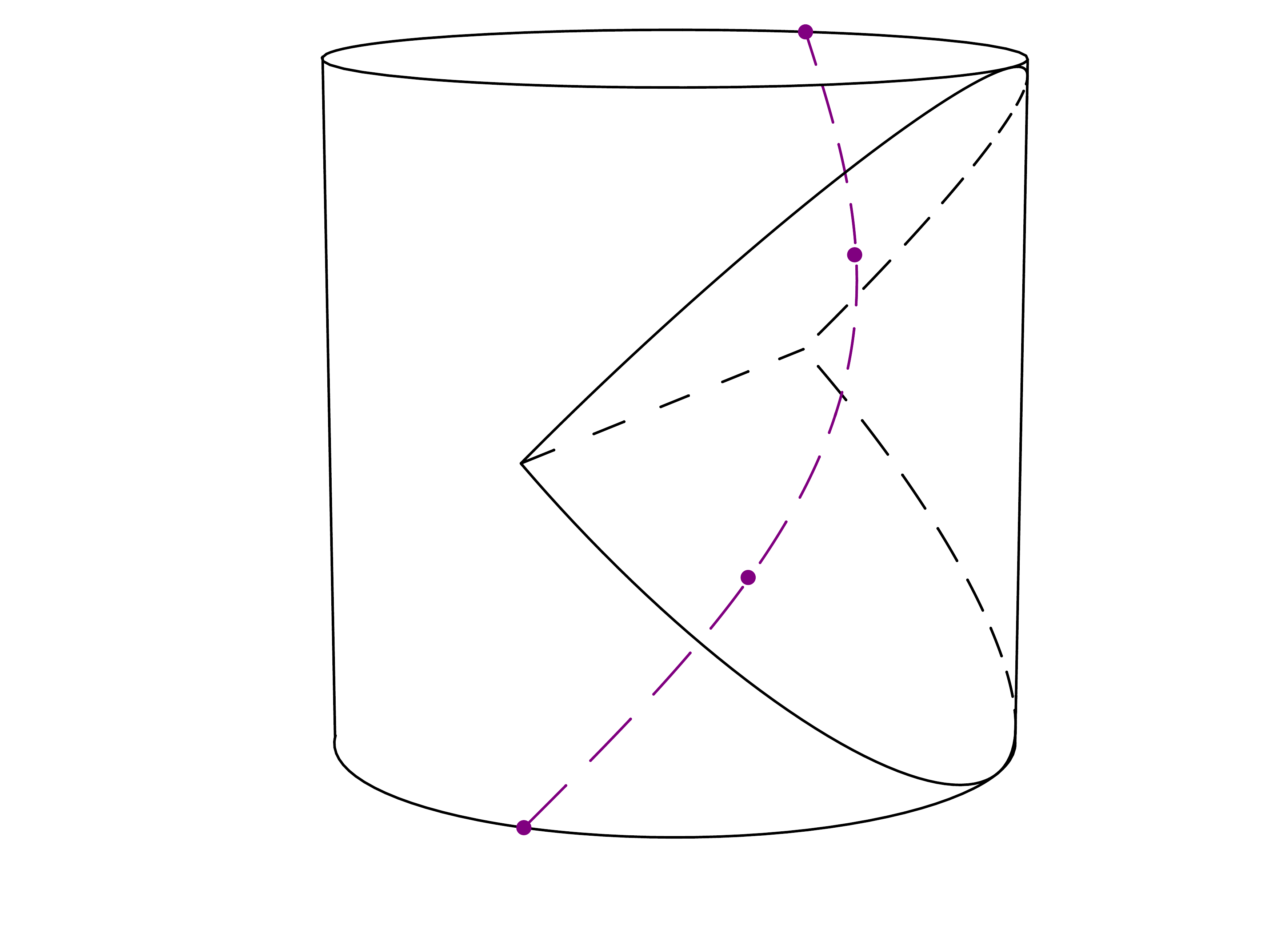}
\caption{This is one of many null geodesics which passes through AdS-Rindler space without reaching the AdS-Rindler boundary. The four highlighted points on the trajectory are (bottom to top) its starting point on the near side of the global boundary, its intersection with the past Rindler horizon, its intersection with the future Rindler horizon, and its endpoint on the  far side of the global boundary.}
\label{fig-nullrindler}
\end{figure}

We propose to extrapolate this condition to AdS and its asymptotic boundary, and subregions thereof. The statement is that continuous reconstruction of a bulk subregion is only possible if every null geodesic in that subregion reaches the asymptotic boundary of that subregion. Applying this to a small diamond on the boundary, we conclude that there is no bulk region for which boundary data on the small diamond can be continuously mapped to a bulk field. As shown in Fig.~\ref{fig-nullrindler}, it is possible to find a null geodesic through any bulk point that does not intersect a small diamond on the AdS boundary. 

A rigorous generalization of the null geodesic criterion to the case of AdS is desirable. In global AdS, at least for the special case of the conformally coupled scalar field, the theorems of Ref.~\cite{BarLeb92} are already strong enough in their current form to ensure continuity. That is because the problem is equivalent to a particular wave equation in a (spatially) compact region with boundary, i.e., the Penrose diagram. And indeed, there we found that the reconstruction was continuous in the way predicted by the theorems.

As stated above, a subtlety arises for the Poincare patch. From the point of view of null geodesics, the Poincare patch is a marginal case. In the Penrose diagram, the boundary of the Poincare patch seems to be just barely large enough to capture all null geodesics passing through the bulk. Why, then, did we find that the reconstruction map is discontinuous, in apparent violation of the theorems of Ref.~\cite{BarLeb92}? In fact, the Poincare patch just barely fails the criterion because the boundary region is not an open set, as required by the theorem that guarantees continuous reconstruction. We believe this may explain the ``infrared" discontinuity we found, and we also believe that a different choice of norm could cure the problem. The existence of an explicit smearing function shows that the problems of the Poincare patch are not fatal.

AdS-Rindler space is of an entirely different character. As we mentioned above, it is clear that there are null geodesics which pass through the bulk and do not intersect even the closure of the boundary. We believe that this is why the discontinuity is in the ``ultraviolet," and also why the smearing function does not exist.

\subsection{Unique Continuation, Null Geodesics,  and RG Flow}

There is another important physical question which brings null geodesics to the fore, and it is less subtle than continuity. The trouble with continuity, as we have seen, is that precise statements depend on a choice of boundary norm, and we have been unable to specify a natural choice for this problem. However, even without a boundary norm, we can ask the bulk question of unique continuation of a solution in the radial direction. In AdS/CFT, the radial evolution of the fields is related to a renormalization group flow of the CFT~\cite{BalKra99,DebVer99,FAuLiu10,HeePol10}. Let $r$ be a radial coordinate such that $r=\infty$ is the boundary, which represents the UV of the CFT. In the CFT, the IR physics is determined by the UV physics, which suggests that a bulk field configuration near $r=\infty$ can be radially evolved inward and determine the field configuration for all $r$. This intuition can be checked for any given  proposed subregion duality.

It is a well-studied problem in mathematics to take a classical field, which solves some wave equation, specified in the region $r>r_*$ and ask if it can be uniquely continued to the region $r<r_*$. If we ask the question locally, meaning that we only ask to continue in a neighborhood of $r=r_*$, then the answer is simple and apparently very robust: the continuation is unique if and only if all null geodesics that intersect the surface  $r=r_*$ enter the known region $r>r_*$. (This is usually stated by saying that the extrinsic curvature tensor of the surface, when contracted with any null vector, should have a certain sign.) The intuition here is the same as with continuous reconstruction: if a null geodesic grazes the surface but does not enter the region where we are given the solution, then we can construct geometric optics type solutions that are zero in the known region, but nonzero inside \cite{Tat04}.

By this same reasoning, one might conclude that reconstruction from the boundary is not unique when there are null geodesics which avoid the boundary, as opposed to the reconstruction being merely discontinuous as stated previously. The resolution has to do with the technical definitions behind the phrasing, which differ slightly between the two questions. In the present context, the non-uniqueness of the solution comes from going all the way to the geometric optics limit along some geodesic which does not enter $r>r_*$. But this is a singular limit, and one might wish to exclude such configurations from being solutions to the equation. That is the choice we implicitly made in previous sections when we talked about continuity. Continuity is broken because of the same type of geometric optics solutions with a singular limit, but we do not have to include the limiting case itself; continuity only depends on the approach to the limit. So the null geodesic criterion, and the reasoning behind it, is the same even though certain technical aspects of the description change based on convenience for the particular question being asked. The point of discussing unique continuation at all is that the boundary is not involved in the question, and so a boundary norm need not be chosen.

For the case of the AdS-Rindler wedge, the same analysis of null geodesics as above indicates that unique continuation fails as well. Knowing the solution for $r > r_*$ does not determine the solution for smaller $r$. Furthermore, the Poincare patch is again a marginal case for this question. Using the standard $z$ coordinate, then for any $z_*$ there are null geodesics which do not deviate from $z=z_*$.

\subsection{The Diagnostic in Other Situations}
To get a sense for how seriously to take our diagnostic, we can apply
it to a variety of familiar situations to test its implications.

\paragraph{AdS black hole formed in a collapse} Suppose we begin
at early times with matter near the AdS boundary, and then at some later time it collapses to make a large black hole. In this case, every null
geodesic reaches the boundary. For a given geodesic, just follow it back in time: at early
times there is no black hole and no singularity, and we know that all
null geodesics in AdS hit the boundary. So for a black hole formed in
a collapse, every null geodesic is captured by the boundary, and it is likely that
continuous reconstruction of the bulk is possible, both inside and
outside the horizon.

\paragraph{Eternal AdS black holes and black branes}
In the case of an eternal black hole, there are some null geodesics that never reach the
boundary; they go from the past singularity to the future singularity.
 The bulk can be continuously reconstructed from the boundary data only outside $r = 3 G_N M$. ($3$ is
the correct numerical factor in 3+1 dimensions. More generally, the bulk
can be reconstructed down to the location of the unstable circular
orbit.)

We will show this explicitly, focusing initially on a spherical black hole in 3+1 dimensions. The metric
is
\be
ds^2 = - f(r) dt^2 + {dr^2 \over f(r)} + r^2 d\Omega_2^2
\ee 
with $f(r) = 1 + r^2/\lads^2 - 2 G_N M / r$. The null geodesics are
extrema of the action
\be
S = \int d \lambda\, g_{\mu \nu} \dot x^\mu \dot x^\nu = \int d \lambda\,
(f \dot t^2 - {\dot r^2 \over f} - r^2 \dot \Omega^2)~.
\ee

\begin{figure}
\centering
\includegraphics[width=6 in]{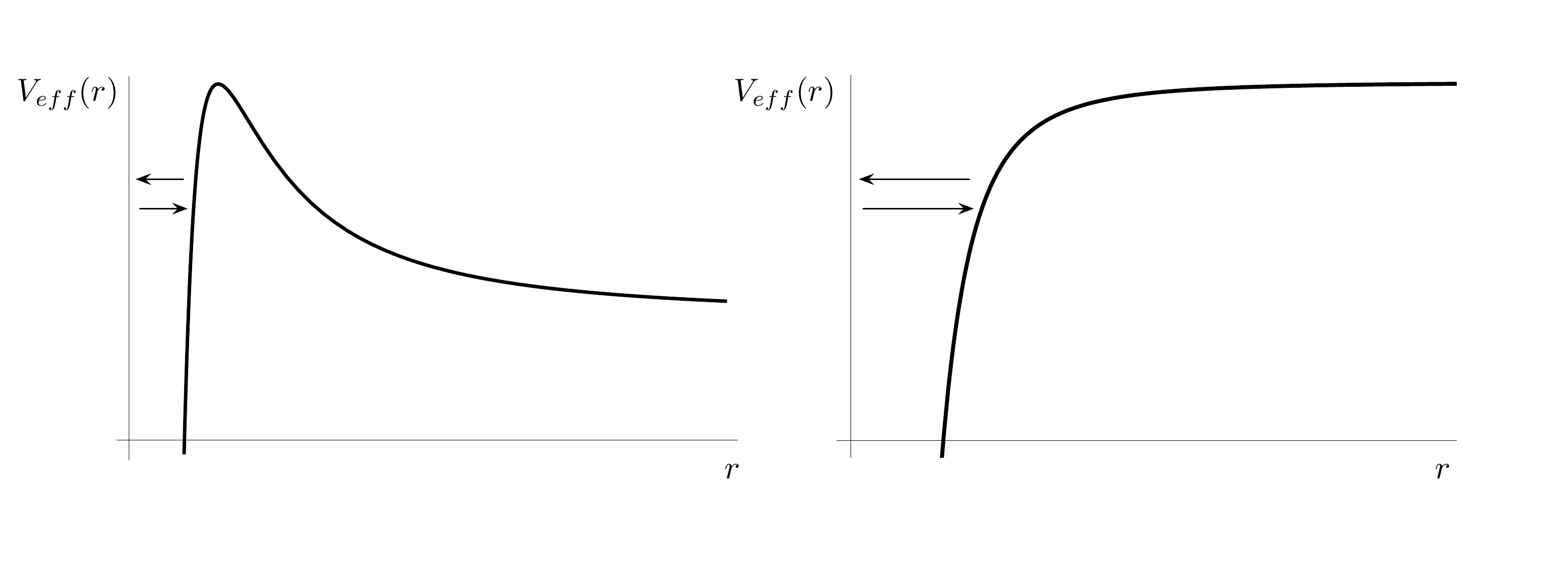}
\caption{On the left we show the effective potential for a null geodesic in a spherical black hole, and on the right the same for a planar black brane. In the case of a spherical black hole, there is a potential barrier which traps some null geodesics in the $r< 3G_NM$ region. Therefore continuous reconstruction from the boundary is not possible for the region $r<3G_NM$. In the planar case, there are null geodesics reach arbitrarily large finite $r$ without making it to the boundary. Hence there is no bulk region which can be continuously reconstructed from the boundary data.}
\label{fig-bhpot}
\end{figure}

Identifying the conserved quantities $E = f \dot t$ and $l = r^2 \dot
\Omega$, the equation of motion can be read off from the
condition that the worldline is null:
\be
0 = g_{\mu \nu} \dot x^\mu \dot x^\nu = {E^2 \over f} - {\dot r^2
  \over f} - {l^2 \over r^2}~.
\ee
This derivation leads to a simple equation for null
geodesics,
\be
\dot r^2 + V_{\rm eff}(r) = E^2 \ \ {\rm with } \  V_{\rm eff} = {f l^2 \over
  r^2}~.
\ee
The effective potential has a maximum at
$r = 3 G_N M$, independent of $l$ (see Fig.~\ref{fig-bhpot}). So null geodesics that begin outside this radius
will inevitably reach the boundary, either in the past or the
future. But there are null geodesics that exit the past horizon,
bounce off the potential barrier, and enter the future
horizon. Because of these, it will be impossible to reconstruct the
bulk region near the horizon.

In this case, rather than conclude that there is anything wrong with
the correspondence, the natural interpretation is that our classical
analysis is breaking down. The ``lost'' null geodesics are being lost
because they fall into the singularity. To recover this information,
we will need to go beyond the classical approximation and resolve the
singularity.

We can also ask about unique continuation. Starting with the data at large $r$, we can try to integrate in to find the solution at smaller $r$. This process will work fine down to $r = 3 G_N M$. However, trying to continue the solution across $3G_NM$ will be impossible.

In the case of a black brane with a planar horizon in ${\rm AdS}_D$, the effective potential for the null geodesics becomes 
\be
V_{\rm eff} = a - {b \over r^{D - 1}}
\ee
where $a$ and $b$ are positive constants. Unlike the spherical black hole, there is no local maximum in the effective potential. For every value of $r$, there are null geodesics which exit the past horizon, travel to that value of $r$, then exit the future horizon. So there is no bulk region that can be continuously reconstructed from the boundary data.

\paragraph{General conclusion about black hole reconstruction}
In the cases of eternal black holes and black branes, the presence of singularities led to the existence of null geodesics which did not reach the boundary, and consequently regions of the bulk which could not be reconstructed from the boundary data. This is not a sign that AdS/CFT is breaking down, but rather an indication that our classical reconstruction procedure is not valid. We know that classical physics breaks down in the neighborhood of the singularity, but the null geodesic criterion suggests that there is a problem even in low-curvature regions. Since the problematic null geodesics begin and end on singularities, it is possible that the physics of singularities needs to be resolved before this question can be answered. A second possibility is that nonlocal boundary operators in the CFT encode the physics of the missing bulk regions. As we emphasize in Section~\ref{sec-duality}, this latter possibility is the expected outcome for AdS-Rindler space, where we believe there is an exact duality between particular bulk and boundary subregions.

\section{Arguments for an AdS-Rindler Subregion Duality}\label{sec-duality}
In this section we exhibit several arguments in favor of a subregion duality for AdS-Rindler space, despite the failure of continuous reconstruction from local boundary fields.

\subsection{Probing the Bulk}

In the previous section, we asked whether we could classically reconstruct the bulk field $\Phi(B)$ from data on the boundary $\phi(b)$. In essence, we restricted ourselves to considering only one point functions $\langle \mathcal{O}(b) \rangle$ on the boundary, and sought to reconstruct bulk fields from integrals of these local boundary operators. However, from an operational standpoint, there is no reason to expect this to be the most efficient way of reconstructing the bulk in general. The boundary theory is equipped with many inherently nonlocal operators. For instance, higher point correlation functions such as $\langle \mathcal{O}(b_1) \mathcal{O}(b_2) \rangle$  could provide a much better probe of the bulk than one point functions.\footnote{It may be the case that higher point functions, which can be obtained by solving classical bulk equations of motion with quantum sources, encounter similar obstructions in the classical limit. However the boundary theory also contains many additional nonlocal operators, such as Wilson loops, which we expect to behave differently in this regime.}

From a physical standpoint, basic properties of AdS/CFT and causality~\cite{SusWit98, BouRan01} are enough to argue that the theory on the boundary diamond should be capable of reconstructing, at the very least, the AdS-Rindler bulk~\cite{BouLei12, CzeKar12, HubRan12}. Consider a bulk observer Bob who lives near the boundary. The boundary theory should be able to describe Bob, and thus it would be inconsistent for Bob to have information about the bulk which the boundary theory does not. Since Bob can send and receive probes into regions of the bulk which are in the intersection of the causal future and causal past of his worldline, he can probe the entire bulk diamond. Thus, it should be the case that the entire bulk diamond can reconstructed from data on the boundary. 

The question of classical reconstruction---restricting to one-point functions on the boundary---amounts to only allowing Bob to make measurements of the field value at his location. If the value of the field decays rapidly near the boundary, Bob would need extremely high resolution to resolve the field. Allowing higher point functions on the boundary amounts to allowing Bob to send and receive probes into the bulk which directly measure the field away from the boundary. This could potentially be a far more efficient way of reconstructing the bulk.

\subsection{Hyperbolic Black Holes}

A CFT dual for AdS-Rindler arises as a special case of the AdS/CFT duality for hyperbolic black holes. The conformal boundary of AdS-Rindler can be viewed as the Rindler patch of Minkowski space, by Eq.~\ref{eq-Rindlermetric2}. The CFT vacuum, when restricted to the Rindler patch, appears as a thermal Unruh state, indicating the presence of a thermal object in the bulk. Indeed, the AdS-Rindler metric in Eq.~\ref{Rindlermetric} with the replacement $z=1/r$ is exactly the $\mu=0$ case of the metric of the hyperbolic black hole studied in Ref.~\cite{Emp99}:
\be
ds^2 = - \left(\frac{r^2}{\lads^2} -1 -\frac{\mu}{r^{d-2}}\right)dt^2 + \left(\frac{r^2}{\lads^2} -1 -\frac{\mu}{r^{d-2}}\right)^{-1}dr^2 +r^2 dH_{d-1}^2~,
\ee
where the spatial hyperbolic plane has the metric
\be
dH_{d-1}^2 = \frac{d\xi^2 + dx_i^2}{\xi^2}~.
\ee

\begin{figure}
\centering
\includegraphics[width=4in]{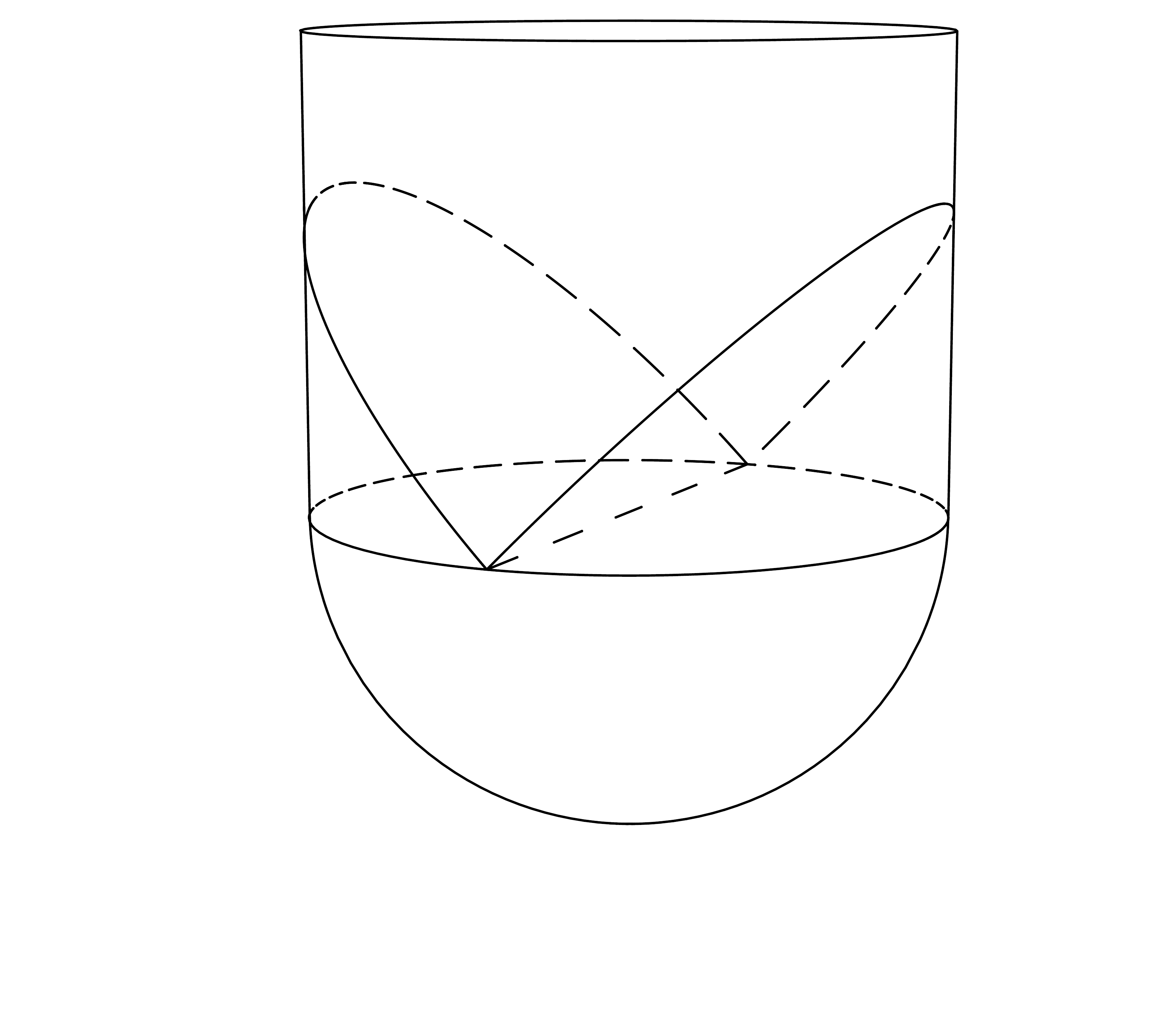}
\caption{The geometry defining the Hartle-Hawking state for AdS-Rindler. Half of the Lorentzian geometry, containing the $t>0$ portion of both the left and right AdS-Rindler spaces, is glued to half of the Euclidean geometry. The left and right sides are linked by the Euclidean geometry, and the result is that the state at $t=0$ is entangled between the two halves.}
\label{fig-HartleHawking}
\end{figure}

\begin{figure}
\centering
\includegraphics[width=4 in]{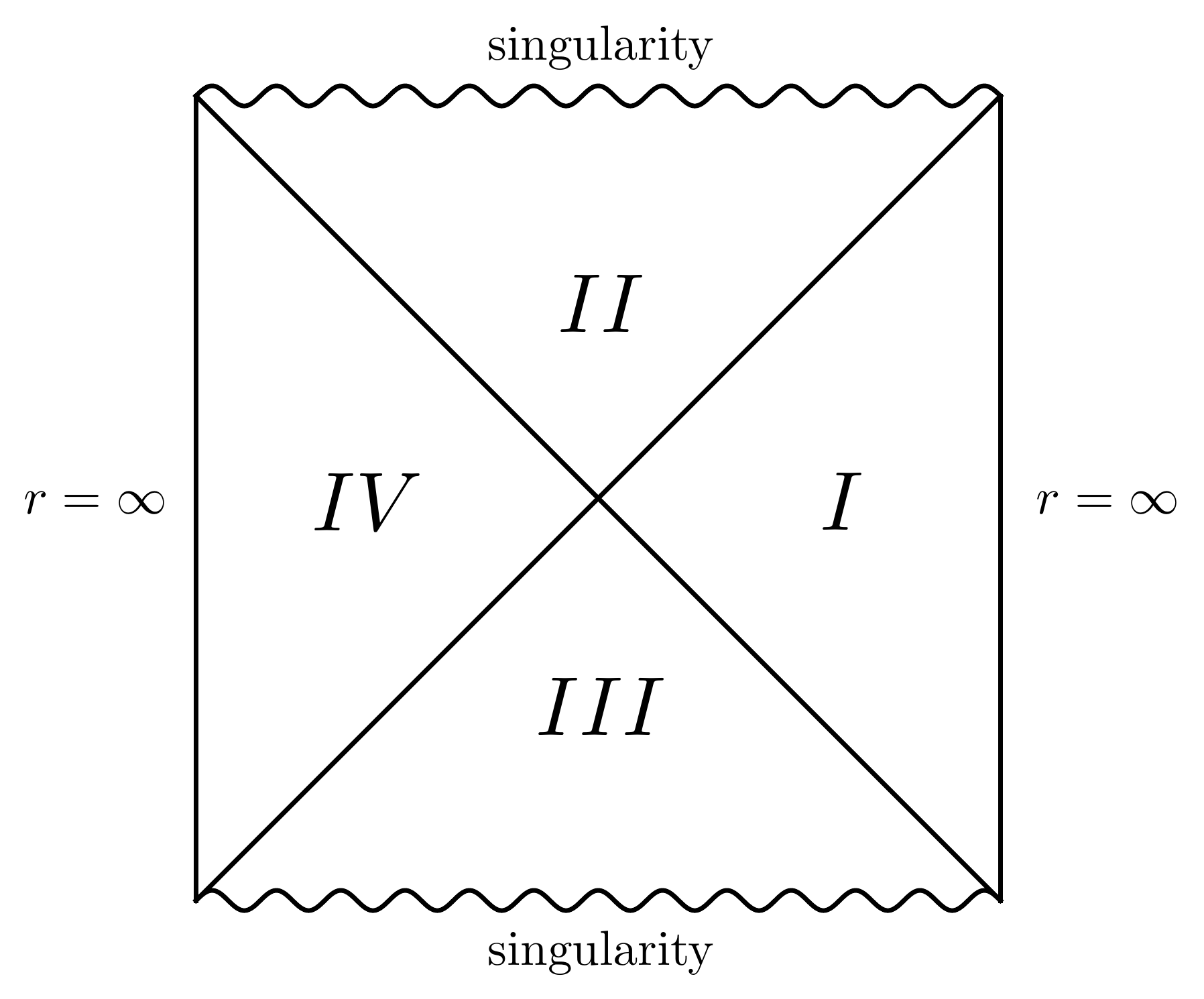}
\caption{The Penrose diagram for a hyperbolic black hole. In the $\mu=0$ case, regions I and IV become the right and left AdS-Rindler wedges.  In this case, the singularity is only a coordinate singularity, so the spacetime can be extended to global AdS.}
\label{fig-hypbh}
\end{figure}

A CFT dual for hyperbolic black holes follows from an adaptation of Maldacena's analysis of the eternal AdS black hole~\cite{Mal01}, which generalizes easily to the hyperbolic case. In particular, hyperbolic black holes have a bifurcate Killing horizon, allowing for a definition of a Hartle-Hawking state from a Euclidean path integral~\cite{Jac94a} (Fig.~\ref{fig-HartleHawking}). The boundary consists of two disconnected copies of ${\mathbf R} \times H^{d-1}$ (the boundary diamonds).  The boundary Hartle-Hawking state is defined through a Euclidean path integral  performed on  $I_{\beta/2} \times H^{d-1}$, where $I_{\beta/2}$ is an interval of length $\beta/2$ and $\beta$ is the inverse Rindler temperature. (Of course, the Hartle-Hawking state for AdS-Rindler is equivalent to the global vacuum. This follows since $I_{\beta/2} \times H^{d-1}$ is conformal to a hemisphere, which is half of the boundary of Euclidean AdS.) The right and left wedges, regions I and IV in Fig.~\ref{fig-hypbh}, are entangled:
\be \label{eq:HH}
\ket{\psi} = \sum_n{e^{- \beta E_n/2} \ket{E_n}_R \ket{E_n}_L}~.
\ee
Restricting to only region I or IV therefore yields a thermal density matrix. 

Excitations above the Hartle-Hawking vacuum can be constructed through operator insertions in the Euclidean geometry. In these states, all particles that enter and leave region I through the Rindler/hyperbolic black hole horizon will be entangled with particles in region IV.  One may therefore question to what extent region I can be reconstructed without access to region IV.  Small excitations above the Hartle-Hawking vacuum, with energy below the temperature $1/2\pi \lads$, will appear as an indiscernible fluctuation in the thermal noise when restricted to region I.  More energetic states, however, are Boltzmann-suppressed.  The density matrix in I will, to a good approximation, accurately register the presence of particles with energy above $1/2\pi \lads$. Hence, the boundary theory of region I, i.e., its density matrix, should encode at least the high-energy states in the bulk region I~\cite{CzeKar12b}.

\section{Discussion}\label{sec-disc}

If an AdS/CFT duality to is to make sense physically, it should be the case that a
physicist with a large but finite computer can simulate the CFT and
learn something about the bulk. Knowing particular boundary
observables to some accuracy should determine the bulk to a
corresponding accuracy. In the case of global AdS/CFT, Hamilton {\it et al.} \cite{HamKab06} found simple
boundary observables---local, gauge invariant operators---which are
sufficient to reconstruct the bulk. We have shown that this reconstruction is continuous, meaning that it is a physical duality in the above sense.

In the case of the proposed AdS-Rindler subregion duality, we have seen that these operators are not
sufficient to perform the same task. We have shown that, given Eq.~\ref{bdryNorm} as our choice of boundary norm, the classical reconstruction map in AdS-Rindler is not continuous. This indicates that we must specify the boundary theory to arbitrary precision to learn anything about the bulk, signaling a breakdown in the physicality of the correspondence.

It is true that our argument for the breakdown depends on the specific boundary norm we choose. We are always free to pick a different norm, for instance one which better respects the symmetries of the boundary theory, and it may be useful to investigate this possibility further. However, the null geodesic criterion gives a simple and intuitive picture of the failure of classical reconstruction, and we would find it surprising if a natural choice of norm could cure the difficulties.

The failure of our diagnostic does not necessarily signal the death of a AdS-Rindler subregion duality. The crucial point is that besides taking the classical limit, we additionally assumed that bulk operators could only be expressed as integrals of local boundary quantities. By removing this extra assumption, a full duality may be recovered---and it would seem surprising if, in general, local boundary quantities were always sufficient for classical reconstruction in all situations. The CFT contains many nonlocal operators, such as complicated superpositions of Wilson loops~\cite{PolSus99,Sus99}, in addition to local ones. Our results suggest that these additional operators are necessary to see locality in the bulk, even in the classical limit.

\acknowledgments
We are grateful to many of our colleagues for extensive and very helpful discussions over the painful course of this work, especially J.~de Boer, V.~Hubeny, J.~Maldacena, D.~Marolf, I.~Rodnianski, S.~Shenker, L.~Susskind, M.~van Raamsdonk, and E.~Verlinde. This work was supported by the Berkeley Center for Theoretical Physics, by the National Science Foundation (award numbers 0855653 
and 0756174), by fqxi grant RFP3-1004, and by the US Department of Energy under Contract DE-AC02-05CH11231. The work of VR and CZ is supported by NSF Graduate Fellowships. The work of SL is supported by a John A. McCone Postdoctoral Fellowship.

\bibliographystyle{utcaps}
\bibliography{all}

\end{document}